\documentstyle[12pt]{article}
\begin{document}
\vspace{0.4cm}
\begin{center}
{\large {\bf CHIRAL SYMMETRY AMPLITUDES IN THE S-WAVE ISOSCALAR AND
ISOVECTOR CHANNELS AND THE $\sigma, f_0(980), a_0 (980)$
SCALAR MESONS}}
\end{center}

\vspace{0.5cm}

\begin{center}
{\large {J.A. Oller and E. Oset}}
\end{center}

\vspace{0.3cm}

{\small {\it
Departamento de F\'{\i}sica Te\'orica and IFIC, Centro Mixto Universidad
de Valencia - CSIC, 46100 Burjassot (Valencia) Spain.}}

\vspace{1cm}

\begin{abstract}
{\small{
We use a nonpertubative approach which combines coupled channel Lippmann
Schwinger equations with meson-meson potentials provided by the lowest
order chiral Lagrangian. By means of one parameter, a cut off in the
momentum of the loop integrals, which results of the order of 1 GeV, we
obtain singularities in the S-wave amplitudes corresponding to the 
$\sigma$, $f_0$
and  $a_0$
resonances.
The  $\pi \pi
\rightarrow \pi \pi \, , \, \pi \pi \rightarrow
K \bar{K}$  phase shifts and inelasticities in the $T = 0$ scalar
channel are well reproduced as well as the $\pi^0 \eta$
and $K \bar{K}$ mass distributions
in the $T = 1$ channel. Furthermore, the total and partial 
decay widths
of the $f_0$  and $a_0$ resonances are properly reproduced including also
the decay
into the $\gamma \gamma$ channel. The results seem to indicate that chiral 
symmetry
constraints at low energy and unitarity in coupled channels is the basic 
information contained in the meson-meson interaction below $\sqrt{s}
\simeq  1.2 \; GeV$.}}
\end{abstract}

\newpage

\section{Introduction}

The understanding of the meson-meson interaction in the scalar sector
is still problematic. There is some debate about the spectrum of hadronic 
states and even more about its nature. Below $\sqrt{s} = 1.2 \;
GeV$, which we will study here,
the existence of a broad scalar-isoscalar meson around $500 \; MeV$ has had
permanent ups and downs. However, the $f_0 (980)$  (also called $S^*$, 
$I^G (J^{PC}) = 0^+ (0^{++}))$,  and the $a_0 (980)$ (also called $\delta$, 
$I^G (J^{PC}) = 1^- (0^{++})) $
mesonic states are well established experimentally, although there is still
debate around their decay widths and particularly about their nature.
Following the discovery of the $f_0$ \cite {1}, and $a_0$ \cite{2}, several
proposals were made about the nature of these states: $q \bar{q}$  states [
3--10], 
multiquark states \cite{11,12} or  $K \bar{K}$ molecules \cite{13,14,15,16}.
Other works argue against the  $q \bar{q}$ nature of the states \cite{17}.
It is
also interesting to investigate the structure of these states in order to
better isolate possible candidates for glueballs and other states rich in
gluons. For instance, in ref. \cite{18} a glueball state at $992 
\; MeV$ ($S_1 (991$)) is
predicted in addition to the $f_0 (980)$
and the $\sigma$ which have another nature. In ref.
\cite{19} a glueball at energies below $0.7 \; GeV$ is 
also predicted which could 
be
a candidate for the $\sigma$. However, its narrow width around $60 \; MeV$ is 
considerably smaller than the $400 \; MeV$ associated to the conventional
$\sigma$ width
in the  $\pi \pi$ interaction. On the other hand in more recent calculations
\cite{20}
the states with a significant component of gluons are only predicted at 
energies around $1.5 \; GeV$ or higher.

The situation in the scalar sector contrasts with the vector and tensor 
sector where the constituent quark models are rather successful in the 
interpretation of the spectrum and properties of the particles.

Coming back to the meson-meson interaction in the scalar sector, a large
fraction of the work done consists in a parametrization of the 
amplitudes respecting general principles \cite{4,5,6,7,21} while some other 
works use models inspired in QCD \cite{11,12,13,14,15} or phenomenological 
ones based on the exchange of mesons \cite{16}.

Along the lines of the amplitude parametrization it is interesting to 
quote the work of \cite{22,23} where it is shown that one pole close to a 
strong threshold ($f.i.$ \, $K \bar{K}$) 
in the II Riemann sheet (we follow the 
notations of ref. \cite{24}) indicates the presence of a molecular state
resulting from the forces between the participant mesons ($K \bar{K}, \pi \pi$
in $T = 0$ or $K \bar{K}, \eta \pi$ in $T = 1$). Conversely, the presence of two
poles close to threshold, 
one in the II sheet
and another one in the IV sheet, would indicate a  $q \bar{q}$ state. The
analysis
in ref. \cite{5,6,7} favours this latter interpretation, but in ref. \cite{21}
the first interpretation is advocated, since the pole found in the IV sheet
is far from threshold. Following this debate, in ref. \cite{7} it is argued 
that the $J/\psi \rightarrow \phi \pi \pi , \phi K \bar{K}$ decay is crucial
in order to distinguish between the two 
interpretations quoted above, questioning the conclusions of \cite{21} and 
offering a good description of these data. However, in ref. \cite{16} the
$f_0$,
appears as a pole in the II sheet and a description of the
$ J/\psi \rightarrow \phi \pi \pi, \phi K \bar{K}$ data 
of the same quality as in \cite{7} is obtained. The main sources of 
discrepancy between \cite{7}  and \cite{21} come from the different ways in
which the background is treated.

Among of the QCD inspired models, in refs. \cite{11,12} the $q^2 \bar{q}^2$ 
states in a bag model is discussed and a rich spectrum of
$SU (3)$
multiplets is obtained which would accommodate the $\sigma, f_0, a_0$ 
mesons in a
same nonet. On the other hand the work in ref. \cite{13,14,15} starts 
from a
colour confining Hamiltonian with hyperfine interaction and studies the same
$q^2 \bar{q}\,^2$ system. According to this latter work the rich 
spectrum predicted in ref.
\cite{11} disappears because the states separate into colour singlets, 
exception made of two weakly bound $K \bar{K}$ states with $T=0$ and $T=1$
which are identified
as the $f_0$ and $a_0$ respectively (the authors also warn about
the influence in 
such states of other meson-meson components due to the coupling of different
channels).

In ref. \cite{8,9} the authors use a unitarized version of the quark model 
and conclude the existence of a $\sigma$ and also that the $f_0, a_0$ are
 manifestations
of $\bar{s} \bar{s}$ and $q \bar{q}$ states respectively, although
they spend most of their time as $K \bar{K}$ components.

In ref. \cite{16} the J\"{u}lich meson exchange model \cite{25} is extended to
account for the meson-meson interaction in the $T=0$ and $T=1$ channels. A
coupled channel approach with the $K \bar{K}$ and $\pi \pi$ channel in $T=0$ 
and 
$K \bar{K}$ and $\pi \eta$ channel  in $T=1$ was
followed and the $f_0$ and $a_0$ states appeared as poles of the t-matrix.

The different approaches which are successful in reproducing the meson-meson
scattering amplitudes rely upon a relatively high number of parameters which
are adjusted to the data, and vary between around 25 in \cite{4,5,6,7,21} and 
5-6 in \cite{8,15}.

One of the properties which has been used to discriminate among models
is the $\gamma \gamma$  decay width of the $f_0$, $a_0$ states \cite{26},
and typically has been
presented as a support for the $K \bar{K}$ molecule nature of the $f_0$ and 
$a_0$ states
\cite{27}. However, the experimental errors are still relatively high and the
data  can be accommodated in several models. 
A brief summary of the present situation in the scalar sector can be
found in ref. \cite{28}.

Another different approach to the meson-meson scattering is chiral 
perturbation theory \cite{29,30,31,32}.

Calculations have been carried out to one loop and one needs counterterms
which require the use of 10 parameters fitted to experimental data.
The approach is valid at energies below $600-700 \,\; MeV$ but such as the 
perturbative calculations are carried out they show obvious limitations to 
face the singularities in the $t$ matrix.

From the former discussion we can see that a key ingredient in most of
the models which successfully describe the amplitudes in the $L=0, T=0,1$
sector 
is the solution of coupled channel scattering equations starting from some 
potential \cite{15,16,33}. On the other hand the success of chiral 
perturbation
theory as an alternative approach to the dynamics of QCD, using mesonic 
instead of  quark degrees of freedom, makes the use of the chiral
Lagrangian extremely appealing. The combination of these two factors is the
main contribution of the present work. In this sense similar steps in the
baryon-meson sector have been done in refs. \cite{34,35} with a remarkable
success. We shall see that this is also the case here. Our model accounts 
automatically for unitary and analyticity and leads to poles in 
the scattering
matrix for the $\sigma, f_0$ and $a_0$ states below $\sqrt{s}= 1.2\, \;  GeV$. 
It predicts the
mass and partial decay widths of the $f_0$ and $a_0$ resonances, as well as
the different scattering amplitudes, in good agreement with experiment and
requires the use of only one parameter which is a cut off in the loop
integrations, following also the spirit of the approach of refs. \cite{34,35}.
This cut off is around $1.2 \; GeV$, very close to the value of the scale of 
chiral symmetry breaking, $\Lambda_{\chi}$ \cite{36}, which plays the similar
role as a scale of energies as our 
cut off.

\section{$L=0, T=0,1$ strong amplitudes in lowest order of chiral
perturbation theory}.

We start from the standard chiral Lagrangian in lowest order of chiral
perturbation theory ($\chi PT$) ${\cal L}_2$ which contains the most general 
low energy
interactions of the pseudoscalar meson octet [29--32]
in this order. This Lagrangian
will provide the potentials which will be used in the coupled channel
scattering equations. 
The interaction chiral Lagrangian is given by

\begin{equation}
{\cal L}_2 = \frac{1}{12 f^2} < ( \partial_\mu
\Phi \Phi - \Phi \partial_\mu \Phi)^2 + M \Phi^4 >
\end{equation}

\noindent
where the symbol $< >$ indicates the trace in the flavour space of the 
$SU(3)$
matrices appearing in $\Phi$ and $M, f$ is the pion decay constant and the
matrices $\Phi$ and $M$ are given by

$$
\Phi \equiv \frac{\vec{\lambda}}{\sqrt{2}} \vec{\phi}= 
\left( \begin{array}{ccc}
\frac{1}{\sqrt{2}} \pi^0 + \frac{1}{\sqrt{6}} \eta_8 & 
\pi^+ & K^+\\
\pi^- & - \frac{1}{\sqrt{2}} \pi^0 + \frac{1}{\sqrt{6}} \eta_8 & K^0\\
K^- & \bar{K}^0 & - \frac{2}{\sqrt{6}} \eta_8
\end{array} \right)
$$

\begin{equation}
M \left(
\begin{array}{ccc}
m_\pi^2 & 0 & 0\\
0 & m_\pi^2 & 0 \\
0 & 0 & 2m_K^2 - m_\pi^2
\end{array} \right)
\end{equation}
\vspace{0.3cm}

\noindent
where in $ M$ we have taken the isospin limit $(m_u = m_d)$. 
From eqs. (1) and (2)
we can write the tree level amplitudes for $K \bar{K} , \pi \pi $ and
$\pi \eta$ (we take $\eta \equiv \eta_8$)

\begin{equation}
\begin{array}{ll}
\hspace{-0.6cm}
\hbox{a)} &  K^+ (k) K^- (p) \rightarrow K^+ (k') K^- (p') \nonumber\\
 & t_a = - \frac{1}{3f^2} (s + t - 2 u + 2 m_K^2)
\end{array}
\end{equation}

\begin{equation}
\hspace{-1.2cm}
\begin{array}{ll}
\hbox{b)} & K^0 (k) \bar{K}^0 (p) \rightarrow K^0 (k') \bar{K}^0 (p') \\
& t_b = t_a
\end{array}
\end{equation}

\begin{equation}
\hspace{-1.1cm}
\begin{array}{ll}
\hbox{c)} & K^+ (k) K^- (p) \rightarrow K^0 (k') \bar{K}^0 (p')\\
& t_c = \frac{1}{2} t_a
\end{array}
\end{equation}

\begin{equation}
\hspace{-0.5cm}
\begin{array}{ll}
\hbox{d)} & K^+ (k) K^- (p) \rightarrow \pi^+ (k') \pi^- (p')\\
& t_d = - \frac{1}{6 f^2} (s + t - 2 u + m_K^2 + m_\pi^2)
\end{array}
\end{equation}

\begin{equation}
\begin{array}{ll}
\hbox{e)} & K^+ (k) K^- (p) \rightarrow \pi^0 (k') \pi^0 (p')\\
& t_e = - \frac{1}{12 f^2} (2s - t -  u + 2 m_K^2 + 2m_\pi^2)
\end{array}
\end{equation}

\begin{equation}
\hspace{-0.6cm}
\begin{array}{ll}
\hbox{f)} & K^0 (k) \bar{K}^0 (p) \rightarrow \pi^+ (k') \pi^- (p')\\
& t_f = - \frac{1}{6 f^2} (s + u - 2t  +  m_K^2 + m_\pi^2)
\end{array}
\end{equation}

\begin{equation}
\hspace{-1.7cm}
\begin{array}{ll}
\hbox{g)} & K^0 (k) \bar{K}^0 (p) \rightarrow \pi^0 (k') \pi^0 (p')\\
& t_g = t_e
\end{array}
\end{equation}

\begin{equation}
\begin{array}{ll}
\hbox{h)} & K^+ (k) \bar{K}^- (p) \rightarrow \pi^0 (k') \eta (p')\\
& t_h = - \frac{\sqrt{3}}{12 f^2} (2s - t - u + \frac{2}{3} m^2_\pi
- \frac{2}{3} m^2_K)
\end{array}
\end{equation}

\begin{equation}
\hspace{-2cm}
\begin{array}{ll}
\hbox{i)} & K^0 (k) \bar{K}^0 (p) \rightarrow \pi^0 (k') \eta (p')\\
& t_i = - t_h
\end{array}
\end{equation}

\begin{equation}
\hspace{-2.5cm}
\begin{array}{ll}
\hbox{j)} & \pi^0 (k) \eta (p) \rightarrow \pi^0 (k') \eta (p')\\
& t_j = - \frac{m^2_\pi}{3f^2}
\end{array}
\end{equation}

\begin{equation}
\hspace{-2cm}
\begin{array}{ll}
\hbox{k)} & \pi^0 (k) \pi^0 (p) \rightarrow \pi^0 (k') \pi^0 (p')\\
& t_k = - \frac{m^2_\pi}{f^2}
\end{array}
\end{equation}

\begin{equation}
\hspace{-1.8cm}
\begin{array}{ll}
\hbox{l)} & \pi^+ (k) \pi^- (p) \rightarrow \pi^0 (k') \pi^0 (p')\\
& t_j = - \frac{1}{3f^2} (2 s - u - t + m^2_\pi)
\end{array}
\end{equation}

\begin{equation}
\hspace{-1.7cm}
\begin{array}{ll}
\hbox{m)} & \pi^+ (k) \pi^- (p) \rightarrow \pi^+ (k') \pi^- (p')\\
& t_m = - \frac{1}{3f^2} (s + t  - 2 u + 2 m^2_\pi)
\end{array}
\end{equation}

\noindent
where $s = (k+p)^2, t=(k-k')^2, u=(k-p')^2$ 
 
In order to obtain the S-wave amplitudes in the different isospin channels
we construct the isospin eigenstates projecting over and S-wave satate
. We have

\begin{equation}
\begin{array}{l}
T = 0\\[3ex]
| K \bar{K} > = - \frac{1}{\sqrt{2}} \sum_{\vec{q}}f(q)
 | K^+ (\vec{q}) K^- (- \vec{q}) + K^0 (\vec{q}) \bar{K}^0 (- \vec{q})>\\[2ex]
| \pi \pi >  = - \frac{1}{\sqrt{6}}\sum_{\vec{q}}f(q)
| \pi ^+ (\vec{q}) \pi^- (- \vec{q}) 
+ \pi^- (\vec{q}) \pi^+  (- \vec{q}) + \pi^0 (\vec{q})  \pi^0 (- \vec{q})>
\end{array}
\end{equation}

\begin{equation}
\hspace{-3cm}
\begin{array}{l}
T = 1\\[3ex]
| K \bar{K} > = - \frac{1}{\sqrt{2}}\sum_{\vec{q}}f(q)
 | K^+ (\vec{q}) K^- (- \vec{q}) - K^0 (\vec{q}) \bar{K}^0 
(- \vec{q})>\\[2ex]
| \pi \eta >  = \sum_{\vec{q}}f(q)|\pi ^0 (\vec{q}) \eta (- \vec{q})>
\end{array}
\end{equation}
\noindent

where $\vec{q}$ is the momentum of the particles in the CM of the pair 
and $q=|\vec{q}|$.
We have used the convention that $| \pi^+ > = - |1,1 >$ 
and $|K^+ > = -|\frac{1}{2}, \frac{1}{2} > $
 isospin states. Note that for 
symmetry reasons there is no $\pi \pi$ S-wave, $T= 1$ state.
The function $f(q)$ is normalized such that $\sum_{\vec{q}}f^2
(q)=1$ and we take it as infinitely peaked around a certain value $q$. 
Note also the apparent extra normalization factor $1/\sqrt{2}$ in the 
$|\pi \pi,T=0>$ state which is a consequence of its symmetry (particles 
and antiparticles go into the same multiplet of isospin). By taking into 
account eqs. (17),(18) and the amplitudes in eqs.
(3) to (15) we can write now:

\begin{equation}
\begin{array}{l}
T = 0\\[2ex]
V_{11} = - < K \bar{K} | {\cal L}_2 | K \bar{K} > = - \frac{1}{4 f^2}
(3 s + 4 m_K^2 - \sum_i p_i^2)\\[2ex]
V_{21} = - < \pi \pi | {\cal L}_2 | K \bar{K} > = - \frac{1}{3 \sqrt{12} f^2}
( \frac{9}{2}s + 3 m_K^2 + 3m_\pi^2 - \frac{3}{2}  \sum_i p_i^2)\\[2ex]
V_{22} = - < \pi \pi | {\cal L}_2 | \pi \pi  > = - \frac{1}{9  f^2}
(9 s + \frac{15  m_\pi^2}{2}  - 3 \sum_i p_i^2) \\[2ex]
\end{array}
\end{equation}

\begin{equation}
\hspace{-0.1cm}
\begin{array}{l}
T = 1\\[2ex]
V_{11} = - < K \bar{K} | {\cal L}_2 | K \bar{K} > = - \frac{1}{12 f^2}
(3 s  - \sum_i p_i^2 + 4 m_K^2)\\[1ex]
V_{21} = - < \pi^0 \eta | {\cal L}_2 | K \bar{K} > = \frac{
\sqrt{3/2}}{12 f^2}
(6 s - 2  \sum_i p_i^2 + \frac{4}{3} m_\pi^2  - \frac{4}{3} m_K^2 )\\[2ex]
V_{22} = - < \pi^0 \eta | {\cal L}_2 | \pi^0 \eta  > = - \frac{1}{3  f^2}
  m_\pi^2 \\[2ex]
\end{array}
\end{equation}

\noindent
where $p_1=k\;,p_2=p\;, p_3=k'\;,p_4=p'$
and the sum over momenta squared runs from 1 to 4. For on shell
amplitudes $p_i^2 = m_i^2.$

\section{$L= 0,  T= 0,1$ strong amplitudes in the coupled channel 
approach}

It is easy to see that if one iterates the lowest order amplitudes 
calculated before, introducing loops, and one evaluates the finite 
contributions to the imaginary part of the amplitudes, the higher order
contributions at energies around 1 GeV or below are even larger than those 
of the lowest order. Furthermore, we have the resonances $f_0$ ,$ a_0$
in the $L=0 $ channel around
the $K \bar{K}$ threshold which should appear as singularities in the $t$
 matrix.
It is then clear that the standard $\chi PT$ expansion, keeping a few orders,
neither will converge nor give this analytical structure. This difficulty
is expectable since one is making an expansion in powers of
 $p^2/\Lambda_\chi^2$, and we
go up to energies $\sqrt{s} = 1.2 \; GeV$.
In order to overcome these problems we assume that the lowest order 
Hamiltonian provides us with the potential that we iterate in the Lippmann 
Schwinger equation with two coupled channels, using relativistic
meson propagators in the intermediate states. This idea follows identical
assumptions made in refs. \cite{34,35} in the meson-baryon sector. Our
channels are labelled 1 for the $K \bar{K}$ and 2 for
 the $\pi \pi$ states in $T=0$
and 1 for $K \bar{K}$, 2 for $\pi \eta$ in  $T = 1$.
The coupled channel equations then become

\begin{equation}
\begin{array}{l}
t_{11}  =  V_{11} + V_{11} G_{11} t_{11} + V_{12} G_{22} t_{21}\\[2ex]
t_{21}  = V_{21} + V_{21} G_{11} t_{11} + V_{22} G_{22} t_{21}\\[2ex]
t_{22}  = V_{22} + V_{21} G_{11} t_{12} + V_{22} G_{22} t_{22}\\[2ex]
\end{array}
\end{equation}

\begin{equation}
G_{ii} = i \frac{1}{q^2 - m_{1i}^2 + i \epsilon} \; \; 
\frac{1}{(P - q)^2 - m_{2i}^2 + i \epsilon}
\end{equation}

\noindent
where $P$ is the total fourmomentum of the meson-meson systems and $q$ the 
fourmomentum of one intermediate meson. The terms VGT in eq. (20)
actually  mean

\begin{equation}
VGT = \int \frac{d^4 q}{(2 \pi)^4} V (k,p;q) G (P,q) t (q;k',p' )
\end{equation}

Note that as $G_{ii}$ has no angular dependence and $V_{ij}$ is purely 
S-wave then $t_{ij}$ is S-wave.  

The diagrammatic meaning of eqs. (20) is shown in fig. 1.

The loop integral in eq. (22) is divergent. In $\chi PT$
 these divergences
are cancelled by counterterms of chiral Lagrangians
at higher order and some finite contribution remains from the loops and
the counterterms. In our approach we take a cut off $q_{max}$
 for the maximum
value of the modulus of the momentum $q$. We vary this parameter in order to
fix some property of the data (for example the position of the maximum
of an experimental cross sections) and after this is done 
there is no freedom in 
our model. 
On grounds of chiral symmetry computations we should expect this cut off
to be around $1 \, GeV$ \cite{36}.

By fixing the cut off one generates the counterterms in our scheme. It is 
also relevant to see that since our scheme produces the
 $\sigma\;, f_0\;, a _0 $
resonances, our work makes connection with the work of ref. \cite{37} where
the finite parts of the counterterms are generated from the exchange of
resonances.

It is also clear that our non perturbative method does not generate all
the terms which appear in the standard $\chi PT$ expansion. For instance loops in
the $t$-channel are not accounted for. But the fact that the scheme proves
so successful, as we shall see, indicates that one is summing the
relevant infinite series needed to produce the structure of the $t$ matrix.
One can expect this because the singularities associated to these  
$t$-channel terms are far away from the physical region, contrary to what
happens with the s-channel terms summed up in the Lippmann Schwinger
equation, which lead to the pole structure of the amplitudes, essential
to reproduce the physical scattering amplitudes.

In principle one would have to solve the integrals in eqs. (20) by taking 
$V$ and $t$ off shell. However, this is not the case, as we show below, at 
least when dealing with S-wave, and we only need the on shell information. 
The argument goes as follows: As we can see from eqs. (18),(19), 
the on shell 
amplitudes are obtained by taking $p_i^2=m_i^2$ and then we can write the 
off shell amplitudes 
as 
$$V=V_{on}+\beta \sum_i (p_i^2-m_i^2)$$
\hspace{0.4cm}
In order to illustrate the procedure let us simplify to one loop and one 
channel (the procedure is easily generalized to two channels). Hence we have

\begin{equation}
V^2=V^2_{on}+2 \beta V_{on} \sum_i (p_i^2-m_i^2)+\beta ^2\sum_{ij}
(p_i^2-m_i^2)(p_j^2-m_j^2)
\end{equation}

When performing the $q^0$ integration in the loop we have two poles, one 
for $q^0=w_1(q)$ and the other one for $q^0=P^0+w_2(q)$, where the indices 
$1$ and $2$ indicate the two mesons inside the loop. Let us take the 
contribution from the first pole (the procedure follows analogously for the 
second pole). From the second term in eq. (23) we get the contribution
\begin{equation}
\frac { 2 \beta V_{on} } { (2 \pi )^3 } \int \frac { d^3q }{ 2w_1(q) } \frac
 { (P^0-w_1)^2-w_2^2 } { (P^0-w_1)^2-w_2^2 }
=\frac { \beta V_{on} } { (2 \pi )^3 } \int dw_1 q
\end{equation}
which for large $\Lambda$ compared to the masses goes as $V_{on} \Lambda ^2$
 and has the same structure in the dynamical variables as the tree 
diagram.
The third term in eq. (23) gives rise to the integral

\begin{equation}
\frac{ \beta ^2 }{ (2\pi )^3 } \int \frac{ d^3q }{ 2w_1(q) }[(P^0
-w_1)^2-w_2^2]=\frac{ \beta ^2 }{ (2 \pi )^3 } 
\int \frac{ d^3q }{ 2w_1(q) }[P^{02}-2w_1P^0+(m_1^2-m_2^2)]
\end{equation}

\hspace{-1.5cm}
coming from the first pole $(q^0=w_1(q))$ and 

\hspace{-1.5cm}
$$\frac{ \beta ^2 }{ (2\pi )^3 } \int \frac{ d^3q }{ 2w_2(q) }
[P^{02}+2w_2P^0+(m_2^2-m_1^2)]$$

coming from the second pole $(q^0=P^0+w_2(q))$. As we can see, the 
terms linear in $P^0$ cancel exactly, respecting the chiral 
structure which does not allow linear terms in $P^0$.

The term proportional to $P^{02}+m_1^2-m_2^2$ in (25) leads again to a 
structure of the type $[P^{02}+(m_1^2-m_2^2)]\Lambda ^2$ and similarly 
happens with the quadratic contribution form the second pole which 
leads to $[P^{02}+(m_2^2-m_1^2)]\Lambda^ 2$. These terms,  
together the $V_{on} \Lambda ^2$ which we obtained before, combine with 
the tree level contribution, giving rise to an amplitude with the same 
structure as the tree level one but with renormalized parameters $f$ and 
masses. However, since we are taking physical values for $f(f=f_\pi=93 \, MeV
)$ and the masses 
in the potential, these terms should be omitted. One can proceed like that 
to higher orders with the same conclusions. 

Since we are taking V and $t$ on shells they  
factorize outside the $q$ integral. Thus the term VGT of eq. (22), after
the $q^0$ integration is performed by choosing the contour in the lower half
of the complex plane, is given by

\begin{equation}
\begin{array}{l}
V_{ij} G_{jj} t_{jk} = V_{ij}
(s) t_{jk} (s) G_{jj} (s)\\[2ex]
G_{jj} (s) = \int_0^{q_{max}} \frac{q^2 dq}{(2 \pi)^2}
\frac{\omega_1 + \omega_2}{ \omega_1 \omega_2 [P^{02} - (
\omega_1 + \omega_2)^2 + i \epsilon]}\\[2ex]
\end{array}
\end{equation}

\noindent
where $\omega_i = (\vec{q}\,^2 + m_i^2)^{1/2}$ 
and $P^{02} = s$
and the subindex $i$ stands for the two intermediate mesons of the $j$
channel.

Thus the coupled channel Lippmann Schwinger equations get reduced to
a set of algebraic equations:

\begin{equation}
At = V
\end{equation}

\noindent
where

\begin{equation}
\begin{array}{l}
t = \left( \begin{array}{c}
t_{11}\\
t_{21}\\
t_{22} \end{array} \right) \quad \quad
V = \left( \begin{array}{c}
V_{11}\\
V_{21}\\
V_{22} \end{array} \right)\\[6ex]
A = \left( \begin{array}{ccc}
1 - V_{11} t_{11} & - V_{12} G_{22} & 0 \\
- V_{21} G_{11} & 1 - V_{22} G_{22} & 0\\
0 & - V_{21} G_{11} & 1 - V_{22} G_{22}
\end{array} \right)
\end{array}
\end{equation}

Introducing the notations

\begin{equation}
\begin{array}{l}
\Delta_\pi = 1 - V_{22} G_{22}\\[2ex]
\Delta_K = 1 - V_{11} G_{11}\\[2ex]
\Delta_c = \Delta_K \Delta_\pi - V_{12}^2 G_{11} G_{22}\\[2ex]
\Delta = det A = \Delta_\pi \Delta_c\\[2ex]
\end{array}
\end{equation}

\noindent
and  inverting the matrix $A$ we obtain the following equations

\begin{equation}
\begin{array}{l}
t_{11} = \frac{1}{\Delta_c} (\Delta_\pi V_{11} + V_{12}^2 G_{22})\\[2ex]
t_{21} = \frac{1}{\Delta_c} (V_{21} G_{11} V_{11} + \Delta_k V_{21})\\[2ex]
t_{22} = \frac{V_{22}}{\Delta_\pi} + \frac{V_{12}^2 G_{11}}{\Delta_\pi
\Delta_c}\\[2ex]
\end{array}
\end{equation}

\noindent
where we have used the fact that $V_{12} = V_{21}$ and $t_{12}=t_{21}$ by time
reversal invariance.

The physical amplitudes as a function of $s$ (real variable) are given by
the expressions in eqs. (30). However, in order to explore the position 
of the poles of the scattering amplitudes one must take into account
the analytical structure of these amplitudes in the different Riemann
sheets. These sheets appear because of the cuts related to the opening
of thresholds in $G_{jj} (s)$  (see fig. 2). By denoting $p_1$ for the 
CM trimomentum
of the $K$ in the $K \bar{K}$ system and $p_2$ for the $\pi$ in the
$\pi \pi, T = 0$ system or the $\pi$ in the 
$\pi \eta,T=1$ systems 

\begin{equation}
p_i = \frac{[s - (m_{1i} + m_{2i})^2]^{1/2}[s - (m_{1i} - m_{2i})^2]^{1/2}}
{2\sqrt{s}}
\end{equation}

The sheets which we consider are

\begin{equation}
\begin{array}{ll}
\hbox{Sheet I}: & Im  \, p_1>0, \, Im  \, p_2 > 0 \\
\hbox{Sheet II}: &  Im  \, p_1>0, \,  Im  \, p_2 < 0 \\
\hbox{Sheet III}: & Im  \, p_1<0, \, Im  \, p_2 < 0 \\
\hbox{Sheet IV:} &  Im \, p_1<0, \, Im  \, p_2 > 0\\[2ex]
\end{array}
\end{equation}

In order to make an analytical extrapolation to the II, III and IV Riemann
sheets we have to cross the cuts $G_{ii}(P^0)$. In order to do this we make use
of the continuity property (for real $P^0 > m_{1i} +
m_{2i})$

\begin{equation}
G_{ii}^{(b)} (P^0 + i \epsilon) = G_{ii}^{(a)} (P^0 - i \epsilon) 
\end{equation}

\noindent
where the index (b) indicates that we are in the second Riemann sheet
of $G_{ii}$ while the index (a) indicates the first Riemann sheet. Then

\begin{equation}
\begin{array}{ll}
G_{ii}^{(b)} (P^0 + i \epsilon) & = 
G_{ii}^{(a)} (P^0 - i \epsilon) =
G_{ii}^{(a)} (P^0 + i \epsilon) - 2 i Im G_{ii}^{(a)} (P^0 + i \epsilon )
=\\[2ex]
& = G_{ii}^{(a)} (P^0 + i \epsilon) + \frac{i}{4 \pi}
\frac{[P^{02} - (m_{1i} + m_{2i})^2]^{1/2} [P^{02} - 
(m_{1i} - m_{2i})^2]^{1/2}}{2 P^{02}}
\end{array}
\end{equation}

We use the first and last terms of these equations to extrapolate 
$G_{ii}^{(b)}$
 in the complex $P^0$ plane (the square root is taken such that Im $\surd >0)$.
Thus the sheet I is obtained with $G_{11}^ {(a)}, G_{22}^ {(a)}$;
the sheet II with $G_{22}^ {(b)}, G_{11}^{(a)}$ ; the sheet III
with $G_{22}^ {(b)}, G_{11}^ {(b)}$ and the sheet IV with $G_{22}^{(a)},
G_{11}^ {(b)}$.

\section{Results}

\subsection{Phase shifts and Inelasticities:}

It is interesting to stress that we have only one free parameter at our
disposal, which for chiral symmetry reasons should be of the the order of the
$1 \; GeV$. 
We have taken as our ultimate choice $q_{max} = 1.1 \; GeV$ which gives a
good agreement with experiment in all channels for phase shifts and decay
widths of the resonances. In order to obtain the phase shifts and
inelasticities we use the two-channel $S$ matrix \cite{15}

\begin{equation}
S = \left[ \begin{array}{ll}
\eta e^{2 i \delta_1} & i (1 - \eta^2)^{1/2} \,  e^{i (\delta_1 + \delta_2)}\\
i (1 - \eta^2)^{1/2} e^{i (\delta_1 + \delta_2)} & \eta e^{2 i \delta_2}
\end{array} \right]
\end{equation}

\noindent
where $\delta_1, \delta_2$ are the phase shifts for the 1 and 2 channels
$(K \bar{K}, \pi\pi $  in $ T =0$ and $K \bar{K},\pi \eta\; $in $ T=1 $ 
respectively) and $\eta$ is the inelasticity. The elements in the $S$ matrix
are related to our amplitudes via:

\begin{equation}
\begin{array}{l}
t_{11} = - \frac{8 \pi \sqrt{s}}{2 i p_1} (S_{11} - 1)\\[2ex]
t_{22} = - \frac{8 \pi \sqrt{s}}{2 i p_2} (S_{22} - 1)\\[2ex]
t_{12} =  t_{21} = - \frac{8 \pi \sqrt{s}}{2 i \sqrt{p_1 p_2}} (S_{12} - 1)\\[2
ex]
\end{array}
\end{equation}

It is interesting to recall that the first two equations allow us to 
determine $\delta_1, \delta_2$ and $\eta$ while the third equation allows
us to determine $\delta_2$  and  $\eta$.
We obtain the same results with both methods, which is a check of consistency 
of the exact unitarity implemented in our scheme.
In fig. 3 we show the results for the $\pi \pi$ phase shifts ($\delta^0_{0\pi
 \pi}$) in $L =0,\;T=0 $. We show in the figure the results
for two values of the cut off energy $\Lambda =
1.1$ and 
 $ 1.2  \; GeV $ ($ \, \Lambda  = (q_{max}^2 + m _K^2) ^{1/2})$

We can see that the agreement with the experiment [1,38--45]
is rather good up to
values of $\sqrt{s}= 1.2 \; GeV$.
From there on discrepancies with experiment start
to appear, but this should be expected because of the opening of the
$\eta \eta $
channel, the increasing importance of the $4 \pi$ channel and the possible 
influence of higher resonances which cannot be considered as 
meson-meson states.
We can see that the position of the $f_0$ resonance is well reproduced. On 
the other hand the results are moderately dependent on the cut off.
 At low
energies, $\sqrt{s} < 500 \, MeV$ we can see that our results are rather
 independent of the
cut off and agree with the data of ref. \cite{45}. Another feature worth
mentioning is a bump in our results around $\sqrt{s} = 500 - 800 \,MeV$
 and in some 
experimental analyses, which in our case is originated, as we shall see,
by the presence of the $\sigma$ pole.

In fig. 4 we show our results for the phase shift of $K \bar{K} \rightarrow
\pi \pi, \, (\delta_1 + \delta_2)$  obtained through
eqs.(35), compared to the experimental results of ref. \cite {46,47}.
The agreement with the data is also good. We have also calculated the 
results with $\Lambda = 1.1 \, GeV$ and  $1.2 \;GeV$ 
and although both are compatible
with the
data, we can  see a better agreement with the data with smaller error
for  $\Lambda =  1.2 \; GeV$. 
The agreement with these two sets of data guarantees agreement with 
the $K \bar{K}$ phase shifts, which are obtained from these  data
by using the two channel unitarity of eq. (35).

In fig. 5 we show the results for the inelasticity using $\Lambda = 1.2
\, GeV$.
There is an obvious discrepancy between different data [45--48].
Nevertheless, we observe a good agreement of our results with the data of
ref. \cite{46} in consistency with the agreement found before for the
phase shifts $\delta^0_{0 \, KK, \pi \pi}$
coming from the same analysis \cite{46}. It should be noted here 
that in ref. \cite{16} one obtains agreement with the upper set of 
data from ref. \cite{48}.

For the $T=1$ channel the data are scarce and different analyses on the
same magnitude, which would show independence on assumptions made, are
lacking. We can compare our results with data for $\pi \eta$ invariant mass
distributions coming from the reaction $ K ^-p \rightarrow \Sigma ^+(1385)
\pi^- \eta $ \cite{49}.
Similarly we can analyse $ K^-p \rightarrow  \Sigma^+ (1385) K^- K^0$.
Following ref. \cite{50} we write

\begin{equation}
\frac{d \sigma_\alpha}{d m_\alpha} =
C | t_\alpha|^2 q_\alpha
\end{equation}

\noindent
where $m_\alpha$ is the invariant mass of the $\pi^- \eta (K^- K^0)$
system, $q_\alpha$ is the $\pi^- (K^-)$
momentum in the $\pi^- \eta  (K^- K^0)$ CM frame, $t_\alpha$ 
is the scattering amplitude of the
$\pi^- \eta (K^- K^0)$ system and $C$ a normalization constant.

In fig. 6 we can see the $\pi^- \eta$ invariant mass distribution and
our results with two different normalization constants.
In fig. 7 we show the same results for the $K^- K^0$ invariant mass with
the corresponding  normalization constants.

The peak in the $\pi^- \eta$
 mass distribution is due to the dominance of the $a_0$ resonance.
We can see that we also produce  the peak at the same place as the
experiment and reproduce 
qualitatively the data. The agreement found with the $K^- K^0$
 mass distribution
is also qualitatively good.

\subsection{Poles and widths}.

In order to determine the "true" mass and width of the 
$\sigma$, $f_0$ and $a_0$
resonances one has to look for the position of the poles in the physical
amplitudes in the complex $P^0$ plane. In the $T = 0$ channel we find a pole in the
II sheet of the complex plane which corresponds to the $f_0$
resonance at $P^0_{f_0} = (981.4 + i \, 22.4) \; MeV$, 
corresponding to a mass $M_{f_0} = 
981.4 \; MeV$  and a width $\Gamma_{f_0} = 44.8 \; MeV$. This singularity 
results in a peak in the physical amplitudes 
$\pi \pi \rightarrow K \bar{K}$ as one can see in
fig. 8.

We find another pole in the $T = 0$ channel corresponding to the $\sigma$.
Actually we find two poles corresponding to zeros of $\Delta_\pi$ 
and $\Delta_c$ in eq.
(29) which are very close. The zero of $\Delta_\pi $  appears at $
P^0_\sigma = (468.5 + i\, 193.6) \, MeV$
while the zero of $\Delta_c$ appears at $P^0_\sigma = ( 469.5 + i 
\, 178.6) \; MeV$.
 The fact that the
two poles are so close and the associated width so large makes that 
in practice one only sees one bump in
the cross section. In any case the second term of $t_{22}$
in eq. (30)  is very small
in the $\sigma$ region in the physical sheet (sheet I). It should be noted
that the $\sigma$ pole in $\Delta_c$ 
also affects the $t_{11}, t_{21}$ amplitudes, but this has
scarce practical consequences since this appears in the unphysical region
below threshold and the physical region near threshold is dominated by
the $f_0$ pole.

The $\sigma$ that we find is rather broad with $\Gamma _\sigma \simeq 400 \; 
MeV$. We should also note that 
since it comes essentially from the first term of $t_{22}$ in eq. (30), it is 
practically tied exclusively to the pion sector, with practically no 
influence from the $K \bar{K}$ sector. Hence, if we eliminate the $K \bar{K}$
 channel from
our coupled equations the $\sigma$ pole survives.

In the case of the $f_0$ we have also observed that it is dominated by the
$K \bar{K}$ channel,  in the sense that if we eliminate the 
$ \pi \pi$ channel
(setting $V_{12} = 0)$ the $f_0$ pole 
 appears also, now at $P^0_{f_0} = (973.4 + i 0)  \, MeV$. The zero width 
is obvious in 
this case since there is no decay into $\pi \pi$, nor possible $K \bar{K}$
decay because
one is below threshold. In the case when the $\pi \pi$ decay channel is open,
the $K \bar{K}$ channel also opens because the finite $f_0$ width gives phase
space for the decay into $K \bar{K}$ (we will come back to this).

In the $T = 1$ channel we find a pole in the II sheet corresponding to the
 $a_0$
at a position $P^0_{a_0} = (973.2 + i \, 85.0) \; MeV$.
 This corresponds to a width of $\Gamma_{a_0} = 170 \; MeV$.
This pole is much tied to the coupled channel. Indeed, if we eliminate
the $\eta \pi$  channel (setting $V_{12} = 0$) the pole disappears.

In fig. 8 and 9  we show the results for the $|t_{12}|^2$ around the 
$f_0$ and $a_0$
regions respectively. We see there two peaks corresponding to the 
resonance excitation around $\sqrt{s} = 987 \; MeV$
 for the $f_0$ and  $ 982 \; MeV$  for the $a_0$. The 
naive widths that one observes there are around $20 \; MeV $ and $  60 \; MeV$
 respectively.
The position of the peaks and the "widths" do not correspond exactly to
the  numbers of the pole position. This is a well know fact of 
resonances. However, in the present case there is one extra feature, which 
is mostly responsible for the shrinking of the width and change  of position
of the maximum. This was discussed in ref. \cite{50} for the specific case
of the $a_0$, but the arguments hold also for the $f_0$.
This feature is a cusp effect, due to the strong coupling of the 
resonances to the $K \bar{K}$  channel which appears around the pole position.
This makes the  $K \bar{K}$  decay width to increase very fast as the energy
increases, reducing the amplitudes accordingly and producing a shift of
the peak position. 
Large widths for the $a_0$ around $200 \; MeV$ have also been claimed in
former studies \cite{16}.

\section{Strong partial decay widths}

The particular structure of the amplitudes $t_{11},
t_{21}, t_{22}$  of eq.(30) makes that
the $t_{11}$ and $t_{12}$ amplitudes in $T = 0$
have a peak around the $f_0$ pole, while the $\pi \pi$
amplitude combines the $f_0$ pole with a background coming from the terms that 
 give rise to the $\sigma$ pole and one has actually
a minimum around that region. In the case of $T = 1$ all these amplitudes
peak around the $a_0$ pole. In view of this, we take the $t_{11}$ and $
t_{12}$
amplitudes in both cases and parametrize them around the resonance
region in terms of a Breit Wigner expression with energy dependent widths.

\begin{equation}
\begin{array}{l}
t_{11} = \frac{1}{2 M_R} \frac{g^2_{R1} (s)}{
\sqrt{s} - M_R + i \frac{\Gamma_R (s)}{2}}\\[2ex]
t_{21} = \frac{1}{2 M_R} \frac{g_{R1} (s) g_{R2} (s)}{
\sqrt{s} - M_R + i \frac{\Gamma_R (s)}{2}}\\[2ex]
\end{array}
\end{equation}

We shall not need the explicit values of $g_{Ri} (s)$ and $\Gamma_R (s)$
since we can
compute our final expressions in terms of $Im \, t_{j1}$ as shown below.

In order to evaluate the partial decays widths we take into account the
finite width of the resonance and substitute in the standard formula for
the decay width

\begin{equation}
2 \pi \delta (M_\alpha - \omega_1 - \omega_2) \rightarrow 2 Im \;
\frac{1}{M_R - \omega_1 - \omega_2 - i \frac{\Gamma_R}{2}}
\end{equation}

\noindent
in view of which the partial decay widths are give by

\begin{equation}
\Gamma_{R, \alpha} = \frac{1}{16 \pi^2} \int^\infty_{m_1 + m_2} d W
\frac{q}{W^2} \, | g_{R \alpha} (W)|^2 \frac{\Gamma_R}{(M_R - W)^2 +
(\frac{\Gamma_R (W)}{2})^2}
\end{equation}

However in practical terms we write eq. (40) in terms of the amplitudes of 
eq. (36) directly and we find

\begin{equation}
\begin{array}{l}
\Gamma_{R,1} = -\frac{1}{16 \pi^2} \int^\infty_{m_1 + m_2} dW \,
\frac{q}{W^2} 4M_R Im t_{11} \\[2ex]
\Gamma_{R,2} = -\frac{1}{16 \pi^2} \int^\infty_{m'_1 + m'_2} dW \,
\frac{q}{W^2} 4M_R \frac{(Im t_{21})^2}{Im t_{11}} 
\end{array}
\end{equation}

We integrate in W around two widths up and down the peak (if allowed by the
lower limit) where we find that  $\Gamma_{R,1} + \Gamma_{R,2} =
\Gamma_{tot}$,
where $\Gamma_{tot}$
is the width corresponding to the pole of the amplitude evaluated before.

In table I we give the results of the partial widths and the mass of the
$f_0$. Both the pole position and the peak of the $t_{21}$ amplitude are in
agreement with the results of the particle data group \cite{51}. 
As with respect to the width, $\Gamma_{tot}$ is also compatible with most
of the present data as can be seen in ref. \cite{51}. The
ratio $\Gamma_{\pi \pi} / (\Gamma_{\pi \pi} + \Gamma_{K \bar{K}})$ is
also in good agreement with the different measurements, as can be seen 
in  table I.
 Since $K \bar{K}$ branching ratio is obtained by subtraction
from unity of the  $\pi \pi$ branching 
ratio, our results are also compatible with
those quoted in the particle data book.

In table II we give the results for the $a_0$. The mass quoted in the
PDB \cite{51} is in agreement with our results for the peak position of
the $t_{21}$ amplitude. Although the position of the pole of the amplitude 
appears at a different energy. This difference between the pole position and
the peak of physical amplitudes is well known in general \cite{62}.
The width of the resonance is $\Gamma_{tot} = 170 \;MeV$,  
apparently larger than many
of the experimental numbers in \cite{51}. However, we already quoted how the
cusp effect can convert this large width into experimental narrower 
amplitudes \cite{50} and this was indeed the case in our results as shown in
fig. 9. We also quoted our qualitative agreement with other theoretical
results \cite{16} which have $\Gamma_{a_0} =
202 \; MeV$. Another interesting piece of data is
the fraction of $K \bar{K}$  decay into $\eta \pi$. 
This fraction is practically unity in
our approach and compatible with the most recent measurement of ref.
\cite{58}. We should note that the fact that the fraction of $K \bar{K}$ 
to $\eta \pi$ 
decay is large is due to the fact that the $a_0$ width is large, since
for narrow resonances there would be little phase space for $K \bar{K}$ 
decay.

As we see, we can claim a global good agreement with experiment for
masses and partial strong decay widths.

\section{Decay of the resonances into $\gamma \gamma$}

In order to determine the decay width of the $a_0$ and $f_0$
 resonances into the
$\gamma \gamma$ channel we proceed in an analogous way as done for the strong decays.
In the first place we obtain the 
$K \bar{K} \rightarrow \gamma \gamma$
 amplitude using the coupled channel
approach discussed above, and second we obtain the effective coupling of
the resonances to the $\gamma \gamma$ channel in order to obtain the partial
decay width via eq. (41).
The Lippmann Schwinger equation for the amplitude 
$K \bar{K} \rightarrow \gamma \gamma$
 would now be written as

\begin{equation}
t_{\gamma 1} = V_{\gamma 1} + V_{\gamma 1} G_{11} t_{11} + V_{\gamma 2}
G_{22} t_{21}
\end{equation}

\noindent
where now $V_{\gamma 1}, V_{\gamma 2}$
 are the potentials obtained from the chiral
Lagrangian at lowest order with external electromagnetic currents
\cite{63,64}:

\begin{equation}
\begin{array}{l}
V_{\gamma \gamma K^+  K^-} = - 2 i e^2 \left\{
\epsilon^1_\mu \epsilon^{\mu (2)} - \frac{k_+ \cdot 
 \epsilon_1 \, k_-  \cdot \epsilon_2}
{k_+ \cdot k_1} - \frac{k_+ \cdot \epsilon_2 \, k_- \cdot
\epsilon_1}{k_+ \cdot k_2}
\right\} \\[2ex]
V_{\gamma \gamma  \pi^0 \eta} = 0\\[2ex]
V_{\gamma \gamma  \pi^+ \pi^-} = - 2 i e^2 \left\{
\epsilon^{(1)}_\mu \epsilon^{\mu (2)} - \frac{p_+ 
\cdot \epsilon_1 \, p_-  \cdot \epsilon_2}
{p_+ \cdot k_1} - \frac{p_+ \cdot 
\epsilon_2 \, p_- \cdot \epsilon_1}{p_+ \cdot k_2}
\right\}
\end{array}
\end{equation}

\noindent
\hspace{.46cm}
In eq. (43), following the nomenclature of ref. \cite{64} we have 
$k_1, k_2$
the photon momenta, $\epsilon_1, \epsilon_2$
 their polarization vectors, and $k_+, k_-, p_+ , p_-$
the momenta of the
$K^+, K^-, \pi^+, \pi^-$ respectively. 
 We also take the same gauge as in ref. \cite{64} in
order to perform the calculations, $\epsilon_1 \cdot k_1 = \epsilon_1
 \cdot k_2 = \epsilon_2 \cdot  k_1 = \epsilon_2 \cdot  k_2 = 0$.
From eqs. (43) and the isospin states of eqs. (16),(17) we
obtain the S-wave $V_{\gamma  i}$ potentials. We factorize the
on shell $t$ matrices and $V_{\gamma 1}$ outside the integrals in 
$V_{\gamma 1} G_{ii} T_{11}$
in eq. (41),
as we did in the strong amplitudes, and we evaluate the integrals of $G_{ii}$
cutting the integral over momentum at the same value of $q_{max}$. 
Since the
scattering amplitudes $t_{i1}$ are evaluated before, the evaluation of 
$t_{\gamma 1}$ 
of eq. (42) is straightforward. Since both $t_{11}$ and $t_{21}$ 
become singular
at the $f_0 , a_0$ poles, the $t_{\gamma 1}$
 amplitude also has a singularity there 
in the $ T = 0$, $T = 1$ channels respectively.

It is interesting to recall than the standard $\chi PT$ expansion acounts 
for some terms not accounted for here \cite{64}. Once again, if these 
terms were added to our amplitude they could not change the resonant part
of the $t_{\gamma 1}$ amplitude of eq.(42) given by 
$t_{11}$ and $t_{21}$ and which is observed
experimentally for $\gamma \gamma \rightarrow \pi^+ \pi^-,
\pi^0 \pi^0, \pi^0 \eta$  \cite{26}.

By following similar steps as for the strong decay we can write

\begin{equation}
\Gamma_{R, \gamma \gamma} = - \frac{1}{2} \frac{1}{16 \pi^2}
\int^\infty_{m_1 + m_2} dW \frac{q}{W^2} 4 M_R
\frac{(Im t_{1 \gamma})^2}{Im t_{11}}
\end{equation}

\noindent
where $m_{1}, m_2$  refer now to $2 m_{\pi}, (m_{\pi} +
m_{\eta})\; $ for $T = 0, (T = 1)$
since below this value there is no imaginary part of $Im
T_{1 \gamma}$. The factor 1/2
in eq. (44) is due to identity of the two final photons. As we did 
before, we integrate over $W$ in an interval of $2\Gamma$ above $M_R$ 
and $2\Gamma$ below, if allowed by phase space.

The results for the widths are shown in tables I and II for the $f_0$
and $a_0$ respectively. We see a very good agreement with the PDB averages 
which are compatible with many of the different results of the analyses quoted 
in those tables. In any case the situation is still problematic, as it is 
reflected in table I in the dispersion of the quoted results of the analyses. 
In fact in this sector,$T=0$ and $L=0$, it is not understood why a broad scalar
 structure above $1 \,GeV$ is observed for $\gamma \gamma \rightarrow \pi^+ 
\pi^-$, following ref. [4], and it is not seen in $\gamma \gamma \rightarrow 
\pi^0
\pi^0$. On the other hand for the $T=1,L=0$ channel the problems come from 
the way one isolates the background around the well observed $a_0(980)$ 
and $a_2(1320)$ resonances in the $\gamma \gamma \rightarrow \pi^0\eta$ 
reaction, with its 
influence on the coupling to two photons of the former resonances. All these 
problems are properly reviewed in \cite{26}.

\section{Conclusions}

We have used a non perturbative approach to deal with the meson meson
interaction in the scalar sector at energies below 
$\sqrt{s} \simeq 1.2 \; GeV$, exploiting 
chiral symmetry and unitarity in the coupled channels. The Lippmann
Schwinger equation with coupled channels, using relativistic meson 
propagators, and the lowest order chiral Lagrangians, providing 
the meson-meson potential, are the two ingredients of the theory. 
One cut off in momentum, expected to be of the order of 
$1 \;  GeV$, is also used in order to cut the loop integrals
and it is fine tuned in order to obtain essentially the position of the
poles. This is the only parameter of the theory. The best fit to all the
data is found with $q_{max} \simeq 1.1 \; GeV$, or equivalently $\Lambda =
1.2 \; GeV$, both for the
$T = 0$ and $T = 1$ channels. We find poles for the $f_0$
and $a_0$ resonances, for $T = 0$ and $T = 1$ respectively,
and also for
the $\sigma$ in the $T = 0$
channel which appears as a broad resonance $(\Gamma = 400 \, MeV)$
at about $500\;  MeV$ of energy.

We compute the $T = 0$ phase shifts and inelasticities for $\pi \pi 
\rightarrow \pi \pi,\,  \pi \pi \rightarrow K \bar{K}$
and $K \bar{K} \rightarrow  K \bar{K}$ and they are
 in good agreement with experiment. In the case $T= 1$ we look for
mass distributions of $\pi^0 \eta$  and $K \bar{K}$
 which are also in agreement with poorer 
experimental data.

We also compute the partial decay widths of the $\pi \pi,
K \bar{K}$  and $\pi^0 \eta, K \bar{K}$ 
respectively and the results are in good agreement
with experiment. We have also studied the partial decay mode of the 
resonances into $\gamma
\gamma$ . We show that the $K \bar{K} \rightarrow \gamma \gamma$
 amplitude (as well as
other ones) contains the $f_0$  and $a_0$  poles in the  $ T = 0$ and 
$T = 1$ channels
respectively and we have evaluated the decay width 
of the $a_0$ and $f_0$ resonances
into the $\gamma \gamma$
channel. The results obtained are in good agreement with experiment.

The scheme used has proved to be very successful. The amount of data  
that one can reproduce using one only parameter is really impressive.
We should recall that fits to some of the data studied here required a
large number of parameters. On the other hand it is remarkable to see
that this parameter is actually of the order of magnitude of what we
could expect from former studies of chiral perturbation theory.

The teaching of the present results is that the constraints of chiral
symmetry at low energies and the implementation of unitarity with the
coupled channels, keeping also the real part of the loops, is the basic
information which is contained in the meson-meson interaction in the
scalar sector below $\sqrt{s} =1.2\,  GeV$.

On the other hand it is quite clear that the present scheme has allowed
us to make progress within the lines of chiral dynamics, introducing 
a non perturbative
method which allow us to go to higher 
energies than in $\chi PT$ and at a lower price in the number
of parameters. It becomes appealing to exploit these ideas in other
processes related to the scalar sector or try to extend it to other
sectors in order to test the potentiality of the method and its limits.

\vspace{0.5cm}

\noindent
{\bf Acknowledgements}.

We would like to acknowledge fruitful discussions with our colleagues,
A. Pich, J. Prades, J. Nieves and W. Weise. One of us J.A. O.
would like to acknowledge financial help from the Generalitat Valenciana.

\newpage
\begin{center}
{\small{Table I: $f_0$ Mass and partial widths ($\Lambda = 1.2 \, GeV$)}}
\end{center}

\begin{center}
\begin{tabular}{|c|c|l|}
\hline
$f_0 $ & our results & $\quad$ experiment\\
\hline
$\begin{array}{c}
\hbox{Mass}\\
\hbox{Pole position}\\
 \, [MeV] \end{array}$  & 981.4 & 980 $\pm$ 10 [51]\\
\cline{1-2}
$\begin{array}{c}
\hbox{Mass} \\
\hbox{Peak of the} \, t_{21}\\ 
\hbox{amplitude}\\
\, [MeV] \end{array}$ & 987 &  980 $\pm$ 10 [51]\\
\hline
$\Gamma_{tot}$ [MeV] & 44.8 & 40 - 100 [51]\\
\hline
$ \frac {\Gamma_{\pi \pi}}{\Gamma_{\pi \pi} + \Gamma_{K \bar{K} } }$ & 0.75 & 
$\begin{array}{l}
0.67  \pm  0.009  \;  [52]\\
0.81 \pm  _{0.009} ^{0.004} \;  [53]\\
0.78  \pm  0.003 \;  [54]\\
\hbox{(av.)} \, 
0.781  \pm  0.024 \;  [51]
\end{array}$\\
\hline
$\begin{array}{c}
\Gamma_{\gamma \gamma}\\
(KeV) \end{array} $ & 0.54 &
$\begin{array}{l}
0.63 \pm 0.14 \, [4]\\
0.42 \pm 0.06 \pm 0.18 \, [55]\\  
0.29 \pm 0.07 \pm 0.12 \, [56]\\  
0.31 \pm 0.14 \pm 0.09 \, [57]\\  
\hbox{(av)} \, 0.56 \pm 0.11  \, [51] \end{array}$\\  
\hline
\end{tabular}
\end{center}

\newpage

\begin{center}
{\small{Table 2: $a_0$ Mass and partial width ($\Lambda = 1.2 \, GeV$)}}
\end{center}

\begin{center}
\begin{tabular}{|c|c|l|}
\hline
$a_0 $ & our results & $\quad \; $experiment\\
\hline
$\begin{array}{c}
\hbox{Mass}\\
\hbox{Pole position}\\
 \, [MeV] \end{array}$  & 937.2 & 983.5 $\pm$ 0.9 [51]\\
\cline{1-2}
$\begin{array}{c}
\hbox{Mass}\\
\hbox{Peak of the} \, t_{21}\\ 
\hbox{amplitude}\\
\, [MeV] \end{array}$ & 982 &  983 $\pm$ 0.9 [51]\\
\hline
$\Gamma_{tot}$ [MeV] & 170 $^{(*)}$ & 50 - 100 [51]\\
\hline
$\frac{\Gamma_{K \bar{K}}}{\Gamma_{\eta \pi}}$ & 1.07 & 
$\begin{array}{l}
1.16  \pm  0.18  \;  [58]\\
0.7 \pm  0.3 \;  [59]\\
0.25 \pm  0.008 \;  [60]\\
\end{array}$\\
\hline
$\begin{array}{cl}
\frac{\Gamma_{\gamma \gamma} \cdot \Gamma_{\eta \pi}}{\Gamma_{tot}}\\
\, [KeV] \end{array} $ & 0.22 & $\begin{array}{l}
0.28  \pm  0.04 \pm 0.01  \;  [55]\\
0.19 \pm  0.07 \pm _{0.1} ^{0.07} \; [61]\\
\hbox{av.} \, 
0.24   \pm _{0.008} ^{0.007}  \;  [51]\\
\end{array}$\\
\hline
\end{tabular}
\end{center}

\noindent 
(*) See discussion in the text about the narrower physical
amplitudes because of the cusp effect.

\newpage
\begin{center}
\bf{FIGURES}
\end{center}

\vspace{0.5cm}
{\bf Fig.1.} Diagrammatic series for eqs. (20) for $t^0_{11}$ with $T=0$

\vspace{0.5cm}
{\bf Fig.2.} Sheets in the complex energy plane. The unitarity cuts are 
indicated by thick lines.

\vspace{0.5cm}
{\bf Fig.3.} $\delta ^0_{0 \pi \pi}$ phase shifts for $\pi \pi \rightarrow
\pi \pi $ in $T=0$ and $L=0$ compared with different analyses indicated 
between brackets.

\vspace{0.5cm}
{\bf Fig.4.} $\delta ^0_{0K \bar{K} \pi \pi}$ phase shifts for $K \bar{K} 
\rightarrow \pi \pi $ in T=0 and L=0 compared with different analyses indicated 
between brackets.

\vspace{0.5cm}
{\bf Fig.5.} $\frac{1-\eta_{00}^{2}}{4}$ in $T=0$ and $L=0$ compared with 
different analyses indicated between brackets.

\vspace{0.5cm}
{\bf Fig.6.} $\pi^-\eta $ mass distributions. The data are from the reaction 
$ K ^-p \rightarrow \Sigma ^+(1385)\pi^- \eta $ [49]. 

\vspace{0.5cm}
{\bf Fig.7.} $K^-K^0 $ mass distributions. The data are from the reaction 
$ K^-p \rightarrow  \Sigma^+ (1385) K^- K^0$ [49]. 

\vspace{0.5cm}
{\bf Fig.8.} $|t_{12}^0|^2$, $T=0$, around the mass of the $f_0 (980)$.

\vspace{0.5cm}
{\bf Fig.9.} $|t_{12}^1|^2$, $T=1$, around the mass of the $a_0 (980)$.

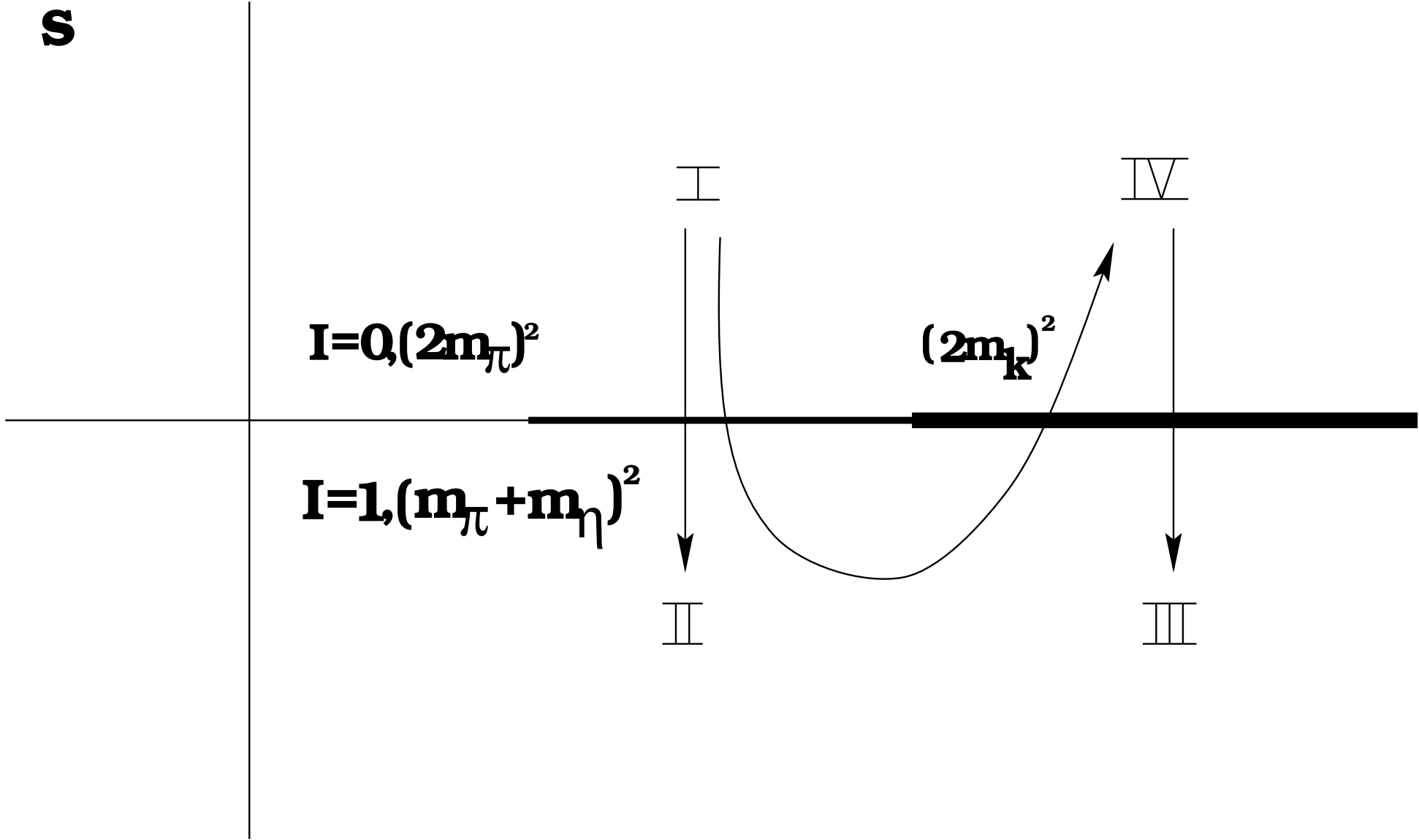
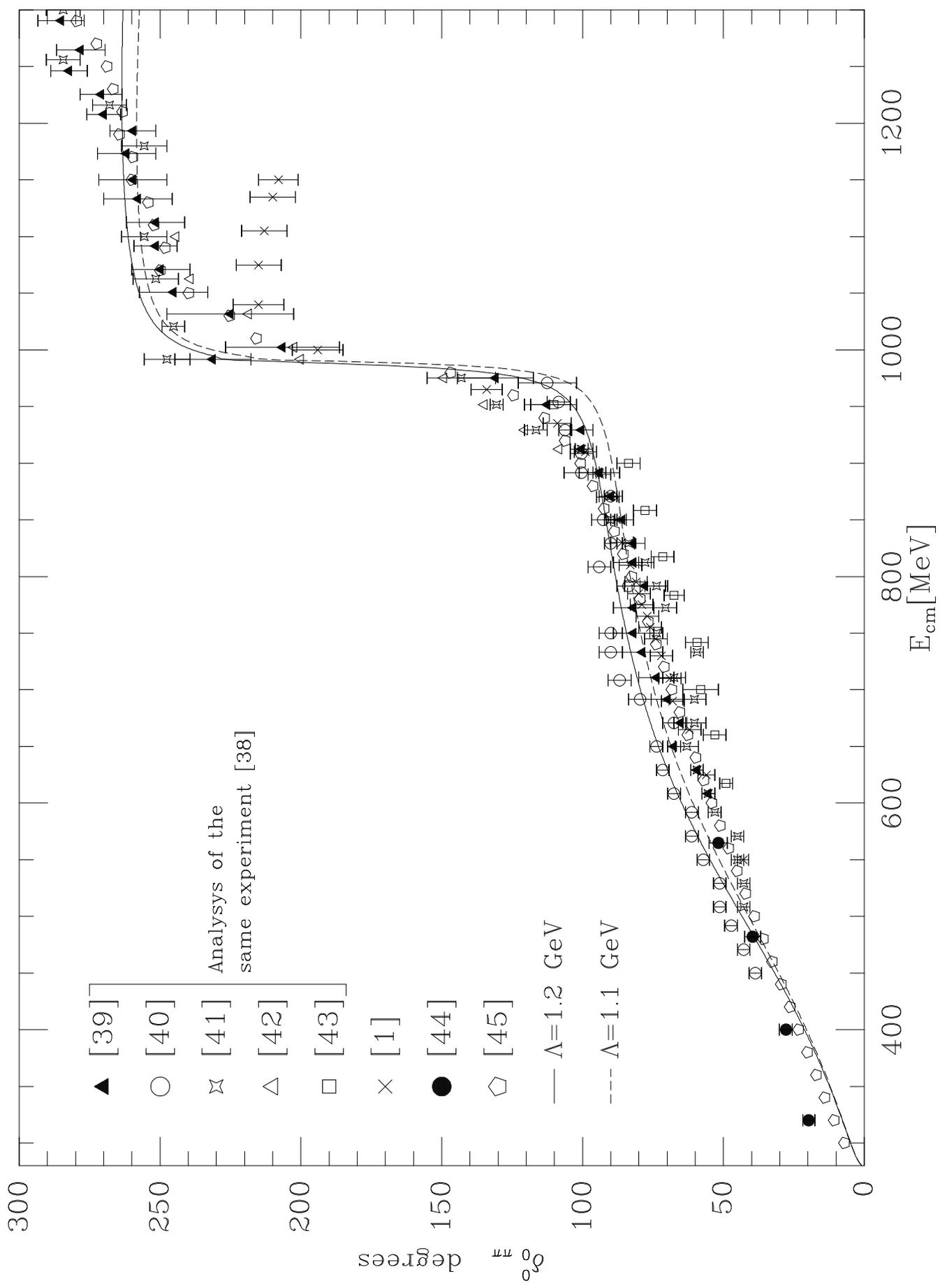
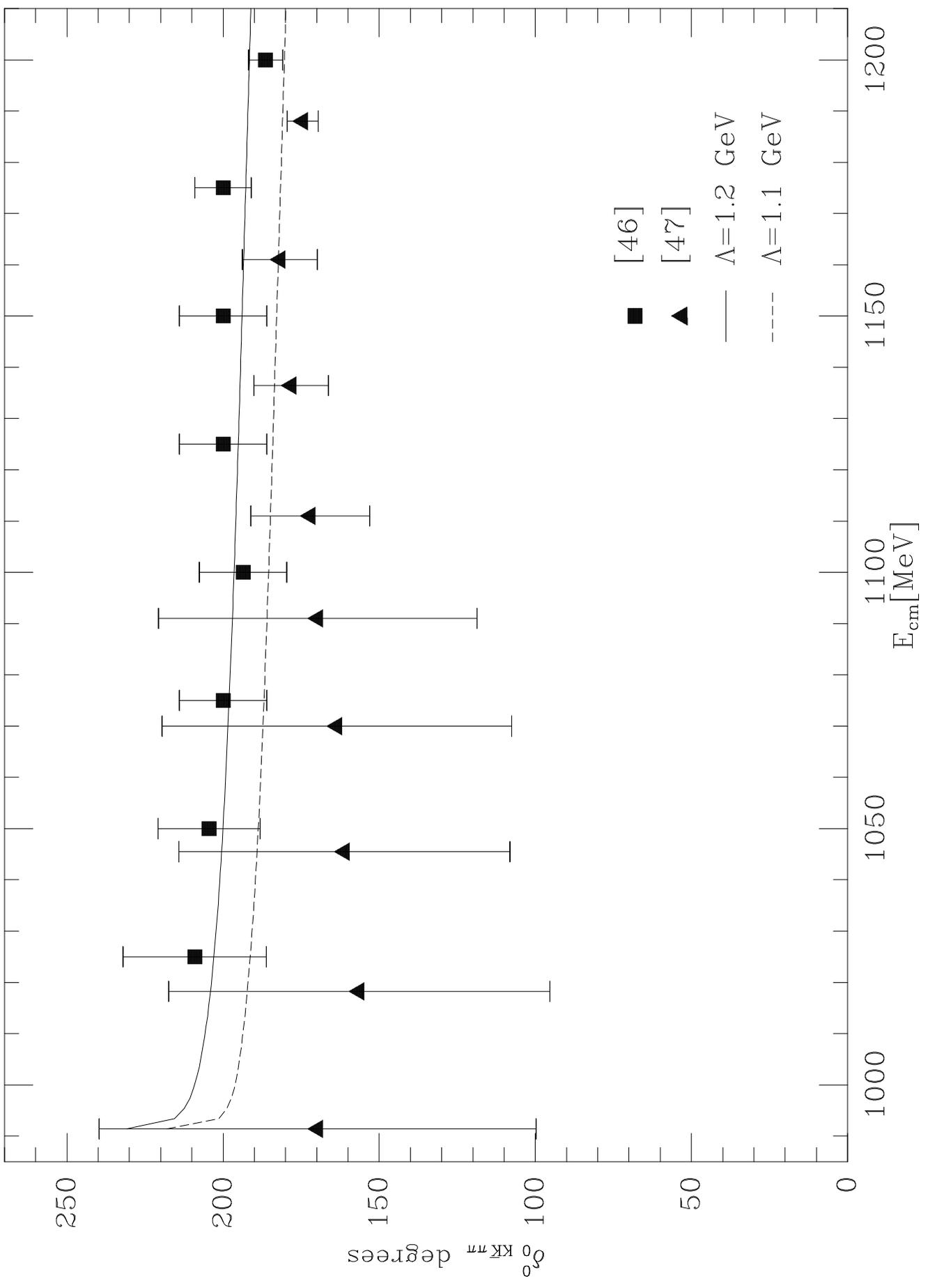
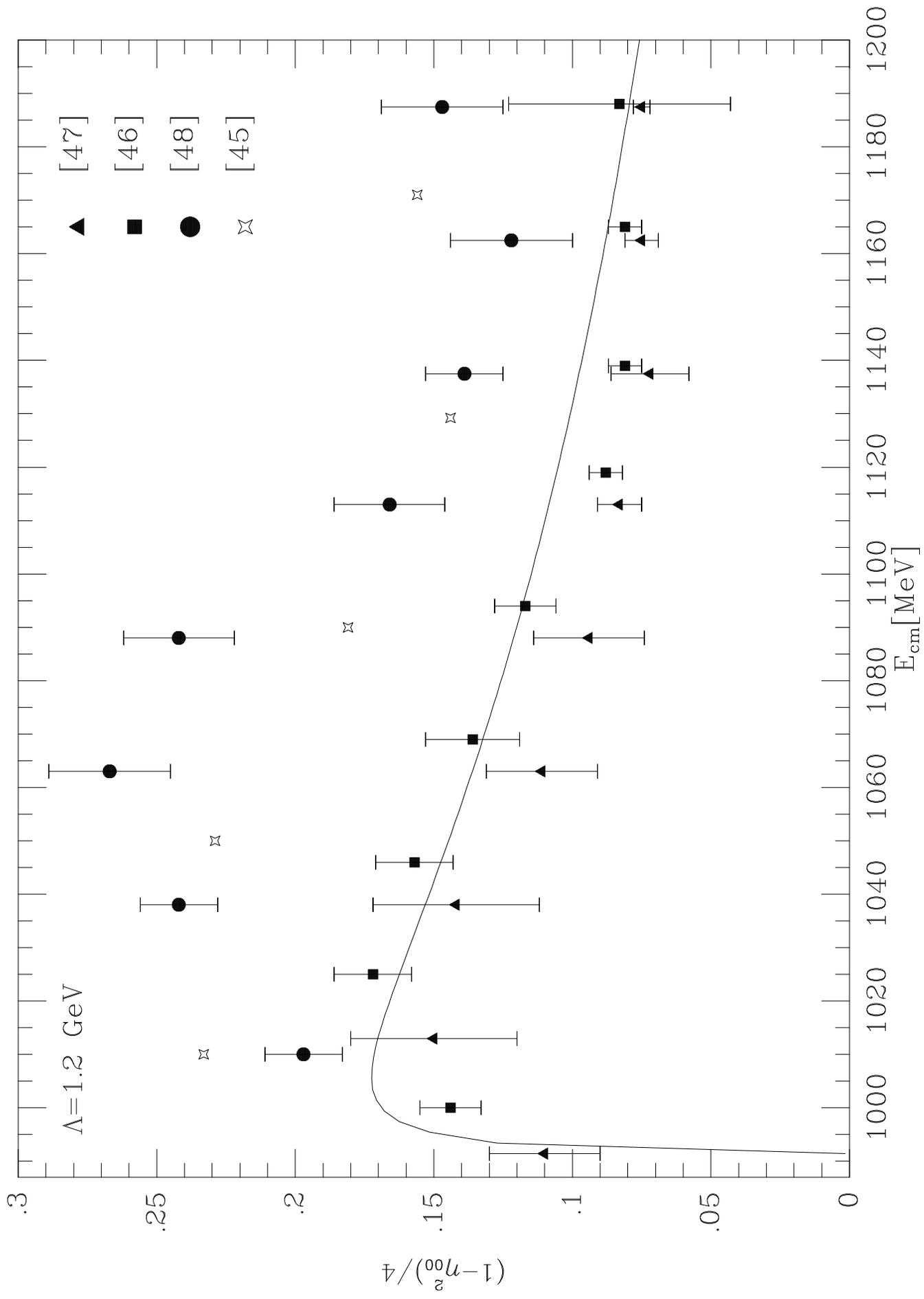
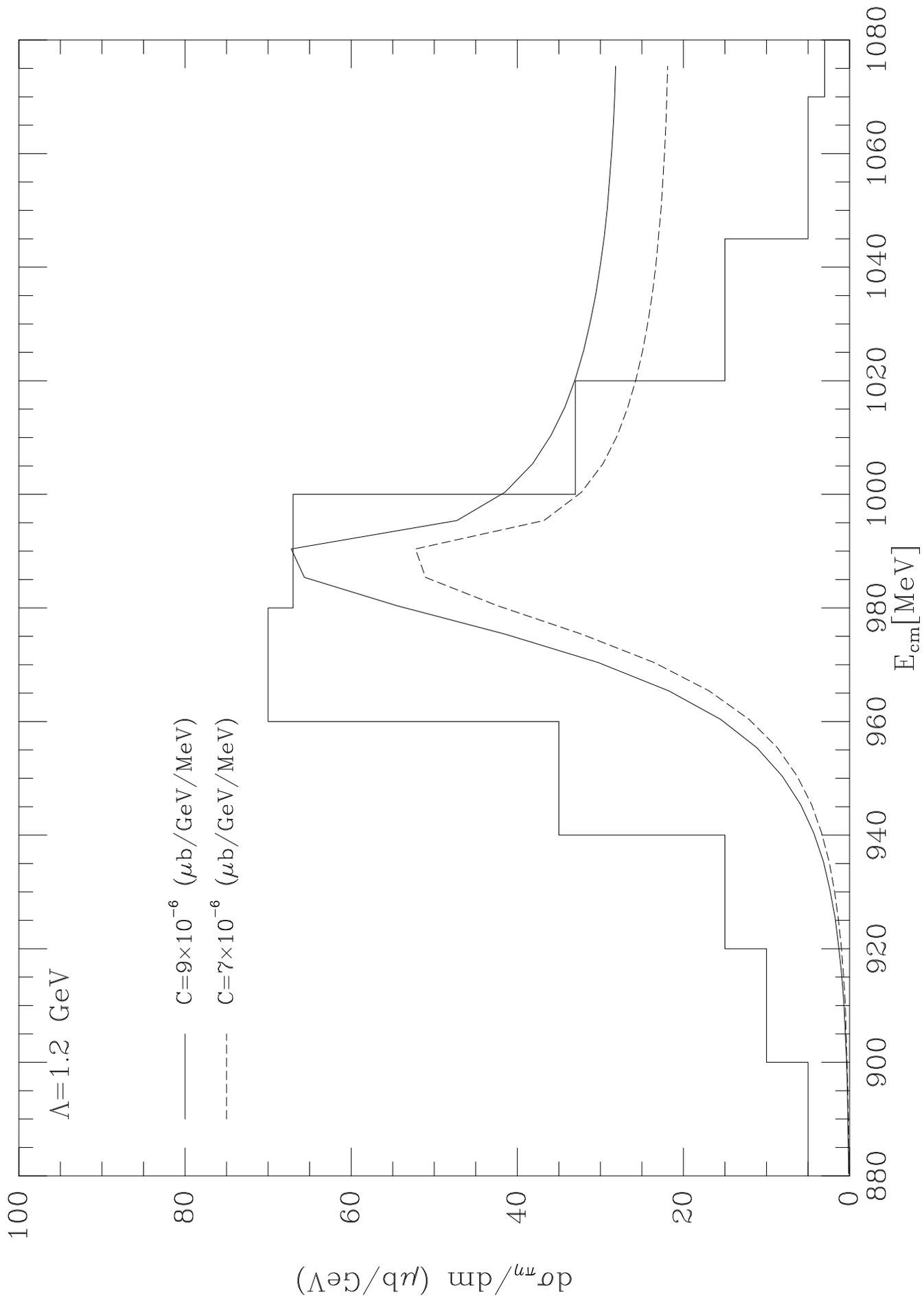
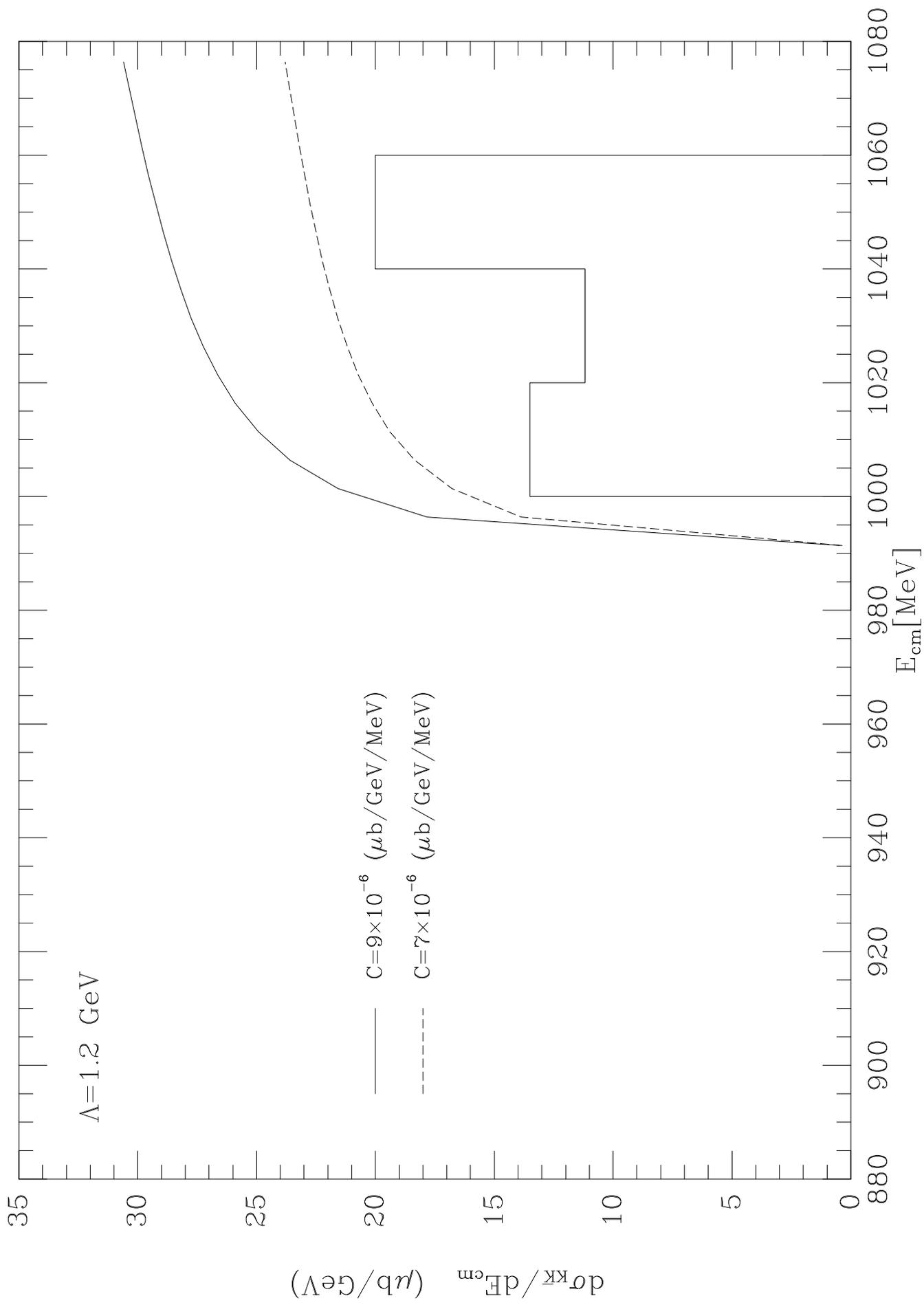
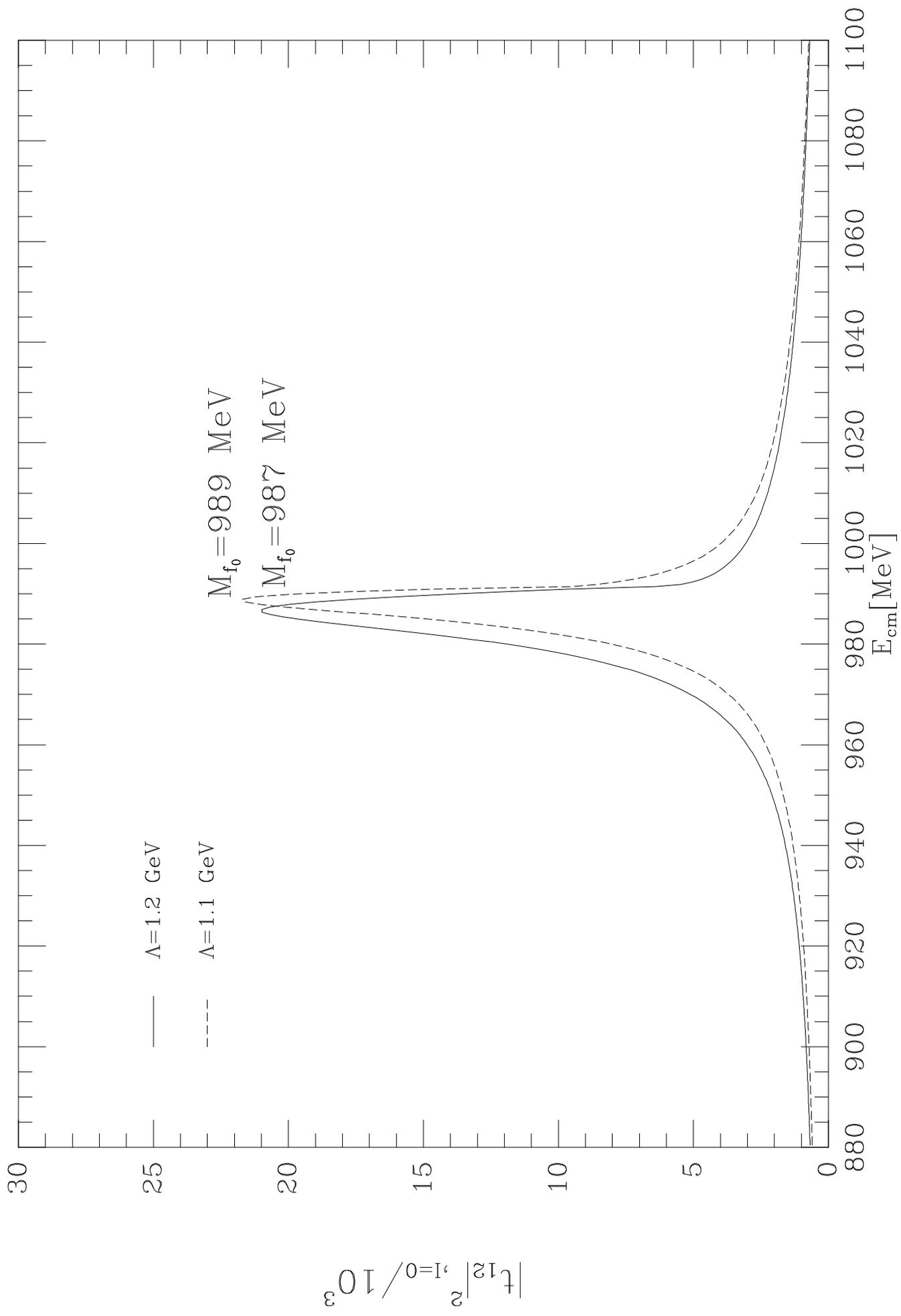
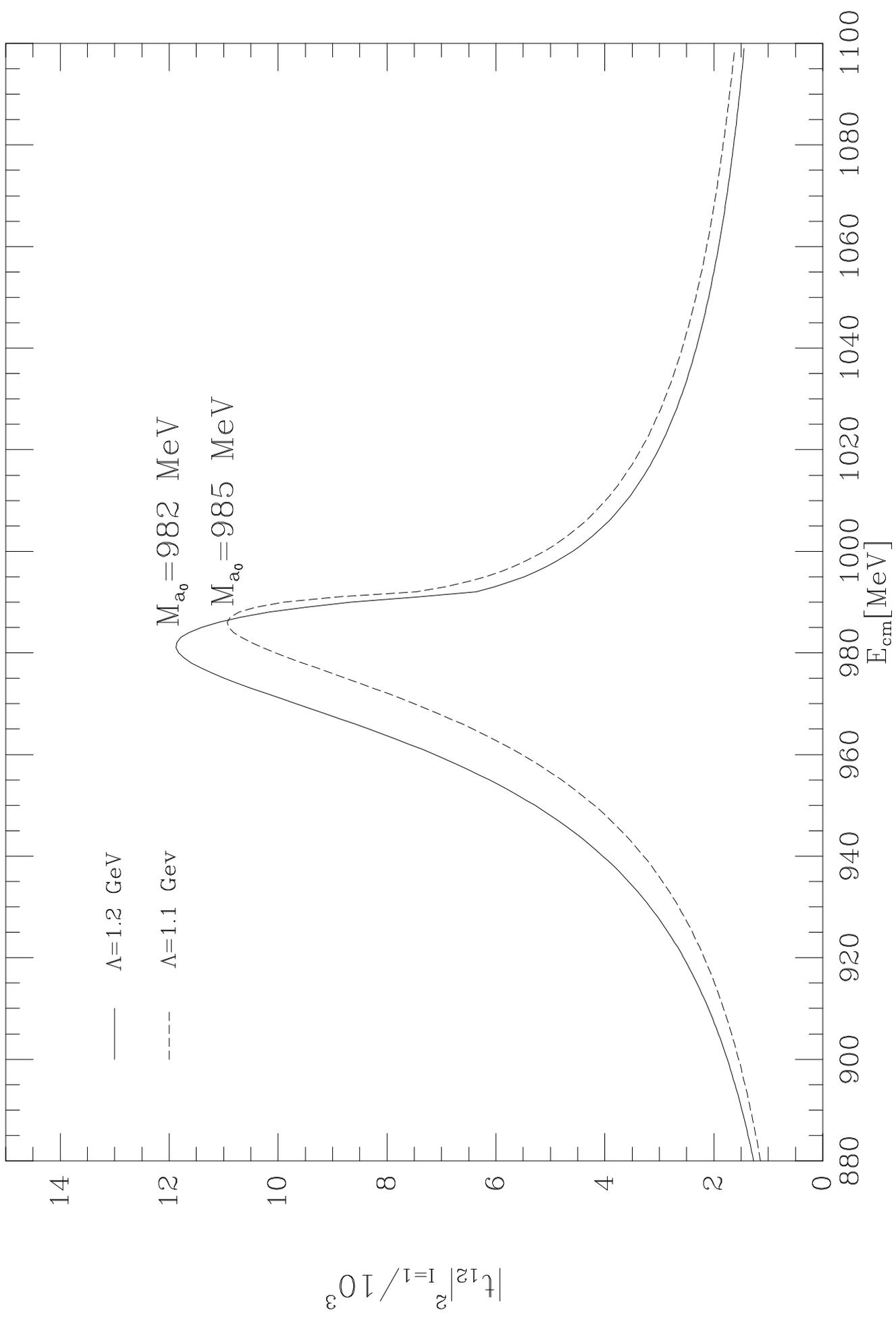
\end{document}